\documentclass[prl,aps,twocolumn,floats,showpacspsfig]{revtex4}
\usepackage{amssymb}
\usepackage{epsfig}

\newcommand{\be}{\begin{equation}}
\newcommand{\ee}{\end{equation}}
\newcommand{\bea}{\begin{eqnarray}}
\newcommand{\eea}{\end{eqnarray}}

\newcommand{\p}{\partial}
\newcommand{\s}{\sigma}

\newcommand{\la}{\langle}
\newcommand{\ra}{\rangle}
\newcommand{\rd}{\mbox{d}}
\newcommand{\ri}{\mbox{i}}
\newcommand{\re}{\mbox{e}}

\newcommand{\vrho}{\mbox{\boldmath $\rho$}}

\begin{document}
\title{Zero energy Majorana modes in spin ladders and a possible realization of the Kitaev model. }

\author{ A. A. Nersesyan$^{1,2}$ and A.  M. Tsvelik$^3$}
{ 
\affiliation{
$^1$ The Abdus Salam International Center for Theoretical Physics, 34100, Trieste, Italy\\
$^2$ Center of Condensed Matter Physics, ITP, Ilia University, 0162, Tbilisi, Georgia\\
$^3$ Department of  Condensed Matter Physics and Materials Science, Brookhaven National Laboratory, Upton, NY 11973-5000, USA}

\begin{abstract}
We show that in double-chain Mott insulators (ladders),
disordered alternating ionic potentials may locally
destroy coherence of magnetic excitations and 
lead to the appearance of spontaneously dimerized islands inside the
Haldane spin-liquid phase.
We argue that a boundary between the dimerized 
and Haldane phases of a spin-1/2 ladder
supports a localized zero-energy Majorana fermion mode. Based on these findings we suggest a realization of a generalized Kitaev model where Majorana fermions can propagate in more than one dimension.  
\end{abstract}


\maketitle

\section{Introduction}
 A possible existence of Majorana zero modes in certain condensed matter systems has been a subject of much discussion in recent literature. The interest is caused by a strong nonlocality of Majorana zero modes which rises hopes of possible applications in such areas as quantum computation \cite{kitaev} and electron teleportation \cite{fu}.  It has been argued that stable Majorana zero energy modes (MZEM) may appear on surfaces \cite{yakovenko} and in  vortex cores of certain superconductors with a nontrivial topological invariant (this includes spinless $p_x + ip_y$ ones)\cite{ivanov}, \cite{roy}. As far as magnetic systems are concerned, the existence of MZEM has been suggested only for a model system  (the Kitaev model)\cite{shankar} which material realization has not yet been achieved. 
 
 This paper presents  two main results. First, we show   that localized MZEM appear on a boundary between spontaneously dimerized (SD) and magnetically disordered phases in systems of two coupled spin-1/2 chains (spin ladders). As was  demonstrated in \cite{shelton},\cite{dim}, the SD phase may exist in spin ladders with sufficiently strong four-spin interaction. There are reasons to believe \cite{lake} that such interactions are present, for instance, in spin ladder material CaCu$_2$O$_3$. Second, we suggest that a balance between competing spin-disordered and SD phases can be shifted by application of a staggered ionic potential which effectively enhances the four-spin interaction. This may be important for applications since realization of such potentials is quite feasible by chemistry means.  Based on these findings we suggest a particular realization of a generalized Kitaev model where MZEM are located on a manifold of arbitrary dimension. As in the original Kitaev model, these dispersionless modes facilitate propagation of  Majorana fermions with nonzero dispersion so that such propagation indeed becomes  becomes  akin to teleportation as described in \cite{fu}. 
 
  The idea that an alternating ionic potential can lead to SD was expressed in \cite{FNG},\cite{FNG1} where the authors studied a transition from the Mott to a conventional ionic band insulator (IBI) in one dimension. It was found that since the Mott and IBI  states differ under a local parity inversion, the transition driven by a change in the strength of the alternating ionic potential proceeds through an intermediate spin liquid phase characterized by a spontaneous dimerization. In the present paper we generalize these ideas for the case of (i) disordered ionic potentials, (ii) the systems consisting of two  half filled chains coupled together. In the second part of the paper we construct a generalized Kitaev model.
 \section{Staggered ionic potential in a single Hubbard chain}
    Let us briefly recall the single Hubbard chain case. At half filling such chain has a spectral gap in the charge sector and gapless excitations in the spin sector. At low energies the spin sector decouples from the charge one and can be considered separetly. For reasons of convenience we will use  the continuous approximation for the entire Hubbard model. This is legitimate if the Mott-Hubbard gap is much smaller than the bandwidth. The effective Euclidian Lagrangian of the Hubbard model in this case is 
\bea
  {\cal L} &=& {\cal L}_c + {\cal L}_s \\
  {\cal L}_c &=& \frac{1}{2K}\Big[v_c^{-1}(\p_{\tau}\Phi_c)^2 + v_c(\p_x\Phi_c)^2\Big] \nonumber\\
  &-& G\cos\sqrt{8\pi}\Phi_c\label{charge}\\
  {\cal L}_s &=& \frac{1}{2}\Big[v_s^{-1}(\p_{\tau}\Phi_s)^2 + v_s(\p_x\Phi_s)^2\Big]
  \label{spin}
 \eea
 where $\Phi_{c,s}$ are charge and spin bosonic fields, $v_{c,s}$ are their velocities, $K$ is the Luttinger parameter and $G$ is the Umklapp coupling constant. All these parameters are related to the bare parameters of the Hubbard model. The cosine term in  (\ref{charge}) is relevant for $K <1$ (repulsion) and opens a gap (Mott-Hubbard gap $\Delta_{\rm MH}$) in the charge sector. The spin sector remains gapless and the correlation functions of the spin fields display power-law decay at $T=0$. 
We omitted in (\ref{spin}) a marginally irrelevant term contributed by
backscattering of the electrons.

 Let us  consider the influence of a staggered ionic potential.  Using the  well known textbook bosonization formulae \cite{textbook2},\cite{textbook} for the staggered charge density   we obtain the following  contribution of such potential to the Lagrangian density:
\bea
&& V = \int \rd x~ V(\pi, x) \rho_{\rm stag}(x)  \\
&& \sim \int  \rd x~ V(\pi,x) \sin\sqrt{2\pi}\Phi_c  \cos\sqrt{2\pi}\Phi_s, \nonumber
\eea
where $V(\pi,x)$ is an envelope made of the Fourier harmonics of the potential concentrated around wave vector $\pi$, and $\rho_{\rm stag}(x)$ is the continuum limit part of the staggered density $(-1)^nC^+_{n\s}C_{n\s}$. Since the charge sector has a gap and the vacuum average of $\sin\sqrt{2\pi}\Phi_c$ is zero,  the effect of
the staggered potential shows up only in the second order of perturbation theory in $V(2k_F=\pi)$ \cite{FNG}:
\bea
\delta H = \lambda\Big[\p_x\Phi_{sR} \p_x \Phi_{sL} - (2\pi \alpha^2)^{-1}\cos\sqrt{8\pi}\Phi_s\Big], \label{marg}
\eea
where $\Phi_{sR;L}$ are chiral components of the scalar field $\Phi_s$,
$\lambda \sim V^2$ and $\alpha$ is a short-distance cutoff of the bosonic theory. More precisely,  the coupling $\lambda$ is expressed as a convolution of  the disorder average of the potentials with the correlation function of the charge fields:
\bea
&& \lambda(x) = \int \rd\tau \rd y ~ V(\pi,x+y)V(\pi,y) \times \label{integ}\\
&& \la\la \sin\sqrt{2\pi}\Phi_c(\tau,x+y)\sin\sqrt{2\pi}\Phi_c(0,y)\ra\ra (\tau^2 + y^2/v_s^2), \nonumber
\eea
 and in principle is coordinate dependent. To provide the convergence of (\ref{integ}) in time domain we need an exponential decay of the dynamical correlation function which occurs only if  the charge gap is nonzero. To estimate the average coupling we may assume that 
 \be
 \la V(\pi,x)V(\pi,0)\ra = V^2\exp(-|x|/\xi).
 \ee
In the most interesting case when the disorder correlation length $\xi \ll v_c/\Delta$ we have $\lambda \sim \frac{\xi \la V^2\ra }{a_0\Delta^2}$. 
  Interaction (\ref{marg}) is marginally relevant and competes with the
   marginally irrelevant interaction of the same form present in  the spin sector of the original Hubbard model.  Therefore such interaction will open a spin gap only when  $\lambda$ exceeds some critical value related to the aforementioned marginally irrelevant backscattering term.  Refs. \cite{FNG},\cite{FNG1}  consider only a uniform potential,
and one could get an impression that the onset of the gapped dimerized phase
would require doubling of the unit cell across the whole chain.
 However, from the above example it follows that, to achieve the desired effect,  it is sufficient to break translational symmetry only locally.

\section{Double chains. The case when charge gaps $>>$ than  the spin ones}

 Now we consider a system of two coupled half-filled chains. For simplicity one may think about them as  Hubbard chains even though  precise details of the charge sector are not  important.  It was shown in \cite{shelton},\cite{dim} that at half filling, when the 
 low-energy dynamics in the spin sector of a single chain coincides with the one  for the  spin S=1/2 Heisenberg model, the continuum limit description of the ladder is given by the model of four Majorana fermions with the Lagrangian density 
\bea
 {\cal L} &=& \frac{1}{2}\chi_R^a(\p_{\tau} - \ri v\p_x)\chi_R^a + \frac{1}{2}\chi_L^a(\p_{\tau} + \ri v\p_x)\chi_L^a  \nonumber\\
&+& \ri m_t\sum_{a=1}^3\chi_R^a\chi_L^a + \ri m_s\chi_R^0\chi_L^0, \label{Maj}
\eea
where the Majorana fields $\chi_R,\chi_L$ in the path integral are real Grassmann numbers and parameters $v, m_t, m_s$ depend on the bare interactions.  This description works well when the interchain exchange is much smaller than the exchange along the ladder. Then parameters $m_t,m_s$ (masses of the triplet and the singlet Majorana fermions) are proportional to the linear combinations of the conventional $J_{\perp}$ and the four-spin $J_{cycle}$ exchange integrals \cite{dim},\cite{gritsev}:
\bea
m_t = AJ_{\perp} + BJ_{\rm cycle}, ~~ m_s = -3AJ_{\perp} + BJ_{\rm cycle}.
\eea
($A,B$ are numerical coefficients). The spectra of the triplet and singlet Majorana fermions are $\epsilon_{t,s}(k) = \sqrt{(vk)^2 + m^2_{t,s}}$; they  do not depend on the signs of $m_{t,s}$. However, since  the spin operators are nonlocal in terms of the fermions, their correlation functions do depend on these signs. In particular, if  $m_tm_s <0$ the ladder is in the spin-disordered (Haldane) state characterized by coherent magnetic excitations with  spectral gap $|m_t|$. If, on the other hand, $m_tm_s >0$ the ground state is spontaneously dimerized (SD) , that is the energy density acquires a nonzero staggered component. Away from quantum critical points  where some of the masses are zero, the spin excitations have spectral gaps but become 
incoherent \cite{dim}.  In most experimental realizations of spin ladders the interchain exchange is either equal or larger than the exchange along the ladder. In that case the singlet gap is large and the corresponding excitation has no influence on the physical properties. However, in such systems as CaCu$_2$O$_3$ where the rung exchange is much smaller than the leg one the situation is different and the singlet mode (though appearing only in the magnetic continua) visibly affects  the dynamical correlation functions \cite{lake}. 

 One may  expect that by arranging a proper coordinate dependence of the interchain interactions one can create a situation when some or all of the Majorana masses change sign. This will lead to creation of Majorana zero modes \cite{jackiw}. In \cite{yulu} it was shown that this does happen when one puts a single vacancy on one of the chains; then all Majorana masses change sign. The result of such change, as it might be expected, is a creation of local spin 1/2 localized at the vacancy. Here we would like to consider a different possibility, namely, when only an odd number of Majorana modes change sign. This happens either on a border between SD and magnetically disordered state or, if the SU(2) symmetry in the spin sector is lost, by a change of sign in one the components of the Majorana triplet. The latter can be achieved by applying an easy axis anisotropy. 
 
 To be definite we will consider the SD case.  We suggest that from the practical point of view a local change of $m_s$ can be achieved by applying staggered ionic potentials acting  on the staggered charge densities of both chains:
 \bea
\delta H = \sum_{i=1,2} V_i(x)(-1)^{n} C^+_{\s,i}(n)C_{\s,i}(n) \label{ionic}
\eea
where $V_i(x)$ is a slow function of $x = na_0$. In order to achieve the desired effect that only $m_s$ changes sign we need $V_1V_2 <0$. In the opposite case $V_1V_2 >0$ the sign change will occur in the triplet mass $m_t$. For simplicity we consider only the former possibiity. To perform actual calculations we adopt  the following hierarchy of energy scales $4t_{\parallel} \gg V, ~
\Delta_{\rm MH} \gg t_{\perp}$, 
 where  $\Delta_{\rm MH}$ is the Mott-Hubbard gap for the individual chains, $V$ is the interchain density-density interaction and $t_{\perp}$ is the interchain tunneling. With such assumptions the high energy degrees of freedom are the charge modes which can be described by the following model: 
\bea
 {\cal L} &=& \frac{1}{2K}\sum_{i=1,2}\Big[v^{-1}_c(\p_{\tau}\Phi_{ci})^2 + v_c (\p_x\Phi_{ci})^2\Big] +  \label{dSG}\\
&+& V\p_x\Phi_{c1}\p_x\Phi_{c2} -  G\left(\cos\sqrt{8\pi}\Phi_{c1}+ 
\cos\sqrt{8\pi}\Phi_{c2}\right), \nonumber
\eea
where fields $\Phi_{c1},\Phi_{c2}$ are  the charge fields of individual chains, $V$  being  proportional to the interchain density-density forward scattering. 
Passing in (\ref{ionic}) to the continuum limit and bosonizing the resulting
expression we obtain:
\bea
\delta H &\sim&  V_1(x)\sin(\sqrt{2\pi}\Phi_{c1})(\epsilon_+ + \epsilon_-)  \nonumber\\
&+& V_2(x)\sin(\sqrt{2\pi}\Phi_{c2})(\epsilon_+ - \epsilon_-),
\eea
where $\epsilon_{\pm}$ are the symmetric and antisymmetric combinations of the chain spin-dimerization operators \cite{def}. In the Majorana representation (\ref{Maj}) these operators are expressed as linear combinations  of products of four order or disorder parameters of the corresponding Ising models (each noninteracting Majorana  fermion is equivalent to the quantum Ising model; see \cite{shelton} and also \cite{textbook2} for details):
\bea
 \epsilon_+ \sim \mu_1\mu_2\mu_3\mu_0, ~~ \epsilon_- \sim \s_1\s_2\s_3\s_0. \label{Eps}
 \eea
 At small distances the following fusion rules take place:
\bea
&& \epsilon^{\pm} \left({\bf R} +  {\vrho}/2\right) \epsilon^{\pm}
\left({\bf R} - {\vrho}/2\right) 
= \label{eps}\\
&& \frac{1}{4}  \left( \frac{\alpha}{\rho} \right) \prod_{i=1}^4 
\left[ 1 \pm \ri \pi \rho~ \kappa_i ({\bf R}) + \cdots   \right]\nonumber\\
&&= {\rm const~} \pm \frac{1}{4}\ri \pi \alpha \sum_{i=1}^4 \kappa_i
- \frac{1}{8} \pi^2 \alpha \rho \left( \sum_{i=1}^4 \kappa_i \right)^2
+ \cdots,
\nonumber
\eea
where $\kappa_i = \chi_R^i\chi_L^i$.  Using fusion rule (\ref{eps}) and keeping only the most relevant terms we obtain in the second order in the potentials the following  shift of all masses:
  \bea
  && m_{s,t }= m_{s,t}^{(0)} + C_{12}(x),\label{masses}\\
  && C_{12}(x) \sim \int \rd y\rd\tau V_1(x+y)V_2(y)\times\nonumber\\
  && ~~\la\la \sin\sqrt{2\pi}\Phi_{c1}(\tau,x+y)\sin\sqrt{2\pi}\Phi_{c2}(0,y)\ra\ra . \nonumber
  \eea
  Naturally, $C_{12}$ vanishes in the absence of interchain interactions. Since in the spin-liquid   state the masses of triplet and singlet Majorana fermions have different signs, such shift may lead to a change in sign of one of the masses. In the uniform case  this would push the system to a spontaneously dimerized phase via Quantum Critical Point \cite{dim}. If the ionic potential is coordinate dependent the sign change will occur locally. For the Hubbard model where the interchain coupling is antiferromagnetic only  the singlet mass changes sign. Then on phase boundaries between SD and disordered phases a single Majorana mode has zero energy states, as it occurs with fermionic states located on the boundary of quasi-1D p-wave superconductor \cite{yakovenko}. Another direct analogy is a boundary between ordered and disordered states in the quantum Ising model. 
  
  A purely one-dimensional case (single ladder) will display disorder effects studied in  \cite{ShTs} and \cite{yulu} where the case of random potential was considered. It was shown that when such boundaries are randomly distributed with a final concentration $n_i$ they give  a singular contrubution to the specific heat: 
  \be
  C(T) \sim n_i\ln^{-3}(T_0/T). \label{CT}
  \ee
 We remark that the models considered in \cite{ShTs},\cite{yulu} dealt with the situation when the number of Majorana zero modes on each domain wall was four.  In that case, as we have already mentioned, each domain wall carries local spin 1/2. 
  
\section{How to construct the Kitaev model}

A much more interesting situation emerges when the number of modes is odd like in the case we consider. Let us consider the above ladder model with large areas of the dimerized phase separated by large areas of the Haldane phase, such that the wave functions of the zero modes from different domain walls do not overlap. Now let us put these ladders on top of a superparamagnet whose  magnetic susceptibility has a peak at $(\pi,0)$ wave vector. So the strongest magnetic fluctuations in such a material are smooth along the ladder direction and oscillate in the direction perpendicular to the ladders. Then they strongly couple  to the 
\be
K^a = \ri(\chi_R^0\chi_R^a + \chi_L^0\chi_L^a), ~~(a=1,2,3), \label{K}
\ee
the  operator representing the difference between magnetization densities of two legs of each ladder \cite{shelton}. Integrating over the bulk magnetic fluctuations  we obtain the effective interaction between   Majorana modes from different ladders:
\bea
&& V_{int} = \frac{1}{2}K^a(y,x)J_{ab}(x-x',r)K^b(y+r,x'), \label{inter}\\
&& J_{ab}(x-x';r) = \la\la N^a(y,x)N^b(y+r,x')\ra\ra, \nonumber
\eea
where $y$ is a location of the ladder in the transverse direction, $x,x'$ are the coordinates along the ladder and $N^a$ is the $a$-th component of the staggered magnetization of the paramagnet.  Since the paramagnet is disordered, this interaction is short ranged. To get the Kitaev model we have to arrange the positions of domain walls in such a pattern  that a wall on one ladder interacts only with one other wall. Such pattern does not need to be regular though disorder will lead to localization of the propagating Majorana modes. 

\begin{figure}
\begin{center}
\epsfxsize=0.3\textwidth
\epsfbox{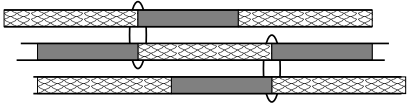}
\end{center}
\caption{Construction of the Kitaev model. Double rails represent spin ladders, the ellipses denote interactions (\ref{inter}) facilitated by the magnetically active media of the substrate described in the main text. Areas with different filling represent SD and Haldane phases. MZEMs are located at boundaries between the phases. }
\end{figure}
As an example we may consider a pattern where each ladder contains alternating SD and Haldane phases of alternating lengths $a$ and $b$. The ladders are shifted by amount $c < a$ in the $y$ direction (see Fig.1).  All lengths are assumed to be much larger than the domain wall width so that the wave functions of MZEMs on different  domain walls do not overlap. Since  the correlation function $J_{ab}$ is short range,  we can neglect all  interactions between domain walls except between the ones located on neighboring ladders. Hence  we get 
\bea
&&H = \label{kitaev}\\
&& \sum_{y,a=1,2,3}\int \rd x \Big[ \frac{\ri v}{2}(-\chi_R^a\p_x\chi_R^a  + \chi_L^a\p_x\chi_L^a) + \ri m_t\chi_R^a\chi_L^a\Big]_y + \nonumber\\
&& \sum_{x_c,y} J_{ab}(y,y+1)\times\nonumber\\
&& (\chi_L^0\chi_L^a + \chi_R^0\chi_R^a)_{y,x_c}(\chi_L^0\chi_L^b + \chi_R^0\chi_R^b)_{y +1,x_c}, \nonumber
\eea
where $x_c$ coordinates mark positions of domain walls. For each wall one has to specify whether it holds a zero mode with right or left chirality and replace $\chi^0$ operator  with its zero mode component: 
\bea
\left(\begin{array}{c}
\chi_R^0(x)\\
\chi_L^0(x)
\end{array}
\right)_y = \frac{\gamma(x_c,y)}{{\cal N}} \re^{ \pm \int_{x_c}^x m_s(x')\rd x'/v} \left(\begin{array}{c}
1\\
\mp 1
\end{array}
\right) + ..., 
\eea 
where the sign depends on the sign of $\delta m_s = m_s(\infty)- m_s(-\infty)$, $\gamma(x_c,y)$ is a zero energy Majorana fermion mode satisfying the Clifford algebra anticommutation relations, ${\cal N}$ is the normalization factor, the dots stand for contributions of higher energy modes. 

Model (\ref{kitaev}) is a particular type of the Kitaev model fermionized by  means of the Jordan-Wigner transformation, as in \cite{feng}.  The latter method, when the original spin model is treated as a model of coupled chains and Jordan-Wigner transformation is done for each chain, represents certain advantages since it allows to avoid a cumbersome gauge fixing used in the original paper \cite{kitaev}. The most important feature of the arrangement depicted on Fig. 1 is that each MZEM in Hamiltonian (\ref{kitaev})   interacts only with one  nearest neighbor exactly as in the original Kitaev model. In this way  the products of neighboring zero Majorana modes $\gamma(x,y)\gamma(x,y+1)$ are integrals of motion and can be replaced by constants. Thus the four fermion term in (\ref{kitaev}) becomes effectively a hopping term for Majorana fermions  $\chi^a$ which now can propagate  in the transverse direction. The latter fermions have spectral gaps, but such sector also exists in the Kitaev model. 

 To make sure that MZEMs are protected, we consider the operators which couple these modes to local perturbations.  The dangerous ones are those which can couple MZEMs located at different positions. Neither  ${\bf K}$ operator considered above nor the staggered components of magnetization can do  this since they contain massive modes and their correlations falls off exponentially. This leaves  the staggered energy density operators $\epsilon_{\pm}$ which are products of four order ( disorder) Ising model operators (\ref{Eps}). The antisymmetric staggered energy density $\epsilon_-$ contains operators $\s_a$ (a=1,2,3) which have no ground state average and therefore vanishes. Operator $\epsilon_+$ has a nonzero ground state average affected by MZEMs and therefore may couple to it. The coupling originates from the change in $\mu_0$;  
  the operators from the triplet sector do not experience any change at the boundary. Using Eqs.(83,84) from \cite{yulu} we get ($x<0$ is in the SD phase):

 \bea
 && \la\epsilon_+(x)\ra = (a_0^4m_t^3m_s)^{1/8} \exp\{- f[m_s(x-x_c)]\}, \\
 && f(x) = \frac{1}{2}\theta(x) x + \frac{1}{8}\Big[K_0(|x|) +  K_{-1}(|x|) \Big], \nonumber
\eea

This function is depicted on Fig.2. Here $m_s$ stands for the asymptotic value of the singlet gap far from the boundary. } As we see, at the boundary where the MZEM is located the symmetric staggered energy density vanishes: $\la\epsilon_+(x)\ra \sim |x-x_c|^{1/8}$. This fact renders the interaction with MZEM  numerically weak. 

\begin{figure}
\begin{center}
\epsfxsize=0.3\textwidth
\epsfbox{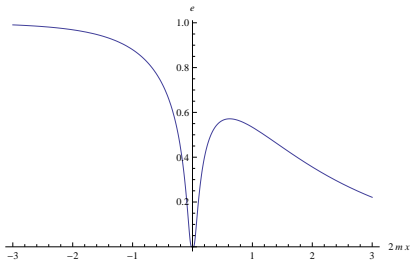}
\end{center}
\caption{$\epsilon_+(x)/\epsilon(-\infty)$ as a function of $2m_s(x-x_c)$.}
\end{figure}

\section{Conclusions}
We have demonstrated that zero energy Majorana modes (MZEMs) robust against external perturbations exist on a boundary between the Haldane and SD phases of spin S=1/2 ladders. We have also suggested a practical way to manufacture  systems with such phase boundaries. Namely, the effective interactions in spin ladders can be manipulated by purely chemical means through staggered ionic potentials. There are two ways MZEMs from different domain walls interact between each other: by direct overlap and via interaction of a particular Fourier component of the magnetization (\ref{K}). The former interaction leads to formation of an impurity band inside of the spin gap with associated singularity in the specific heat (\ref{CT}). However, by maintaining a proper distance between the phase boundaries one can make this overlap small. For this case we have suggested a particular arrangement when the interactions between MZEMs from different ladders create a network similar to the one existing in the Kitaev model in its gapped phase.

   We are   grateful to B. L. Altshuler and V. Gritsev for valuable discussions. A. M. T. is grateful to Abdus Salam ICTP for kind hospitality and  also acknowledges a support from the US DOE, Basic Energy Sciences, Material Sciences and Engineering Division. A. N. N. is supported by grants GNSF-ST09/4-447 and IZ74Z0-128058/1.

\end{document}